\documentclass[3p,times,twocolumn,superscriptaddress,preprintnumbers,nofootinbib,preprint]{elsarticle}

\usepackage{ecrc}

\volume{00}

\firstpage{1}

\journalname{Nuclear Physics B Proceedings Supplement}

\runauth{E.L.Berger and H. Zhang}

\jid{nuphbp}

\jnltitlelogo{Nuclear Physics B Proceedings Supplement}

\usepackage{amssymb}
\usepackage{amsthm}
\usepackage{graphicx}
\usepackage{amsmath}

\biboptions{comma,square}

\usepackage[figuresright]{rotating}

\newcommand{\be}{\begin{equation}}
\newcommand{\ee}{\end{equation}}
\newcommand{\bea}{\begin{eqnarray}}
\newcommand{\eea}{\end{eqnarray}}

\def\met{\slash{\!\!\!\!E}_{\text{T}}}
\begin{document}

\begin{frontmatter}



\dochead{}

\title{Higgs boson physics and broken flavor symmetry -- LHC phenomenology\tnoteref{label1}}
\tnotetext[label1]{Paper presented by E~L~Berger at the 37th Conference on High Energy Physics, July 2 - 9, 
2014, Valencia, Spain.}



\author[label2]{Edmond L. Berger}
%\ead{berger@anl.gov}
\address[label2]{High Energy Physics Division, Argonne National Laboratory, 
Argonne, IL 60439, USA}

\author[label3]{Hao Zhang}
%\ead{zhanghao@physics.ucsb.edu}
\address[label3]{Department of Physics, University of California, 
Santa Barbara, CA 93106, USA}

\begin{abstract}
The LHC implications are presented of a simplified model of broken flavor symmetry in which a new scalar 
(a flavon) emerges with mass in the TeV range.  We summarize the influence of the model on Higgs boson physics,  notably on the production cross section and decay branching fractions.  Limits are obtained on the flavon $\varphi$ 
from heavy Higgs boson searches at the LHC at 7 and 8 TeV.  The branching fractions of the flavon are computed 
as a function of the flavon mass and the Higgs-flavon mixing angle.  We explore possible discovery of the flavon at 14 TeV, particularly via the $\varphi \rightarrow Z^0Z^0$ decay channel in the $2\ell2\ell'$ final state, and through standard model Higgs boson pair production $\varphi \rightarrow hh$ in the $b\bar{b}\gamma\gamma$ 
final state.   The flavon mass range up to $500$ GeV could probed down to quite small values of the Higgs-flavon 
mixing angle with 100 fb$^{-1}$ of integrated luminosity at 14 TeV.  
\end{abstract}

\begin{keyword}
Higgs boson \sep broken flavor symmetry \sep flavon \sep Higgs-flavon mixing \sep LHC phenomenology  


\end{keyword}

\end{frontmatter}

% \linenumbers

\section{Introduction}
\label{sec:intro}
The standard model (SM) of particle physics describes physics at the electroweak symmetry 
breaking (EWSB) scale of the visible sector remarkably well.  Following the discovery of a 
narrow state with mass near $125$~GeV, whose properties are much like those expected of 
the SM Higgs boson, all of the expected SM particles can be said to have been detected.  
Attention has turned to precise determination of the properties of this Higgs boson, notably 
its decay branching fractions, and to a more detailed exploration of physics at the electroweak 
scale, including the search for possible new physics beyond the SM.  

The flavor structure of the SM remains a puzzle, and it is intriguing to consider how broken 
flavor symmetry may be related to broken electroweak symmetry.  The fact that no significant 
deviations from SM predictions have appeared in flavor-related physics processes indicates that 
that any TeV scale new physics (NP) does not introduce an important new source of flavor change 
or CP violation beyond the SM.  This inference hints at flavor symmetry (horizontal symmetry) in a 
NP model. The idea that the NP interactions are invariant under a flavor symmetry group is known 
as minimal flavor violation (MFV)~\cite{D'Ambrosio:2002ex}.  In the MFV scenario, the SM flavor 
symmetry is broken explicitly by the non-vanishing SM Yukawa coupling constants. The flavor symmetry 
could nevertheless be a true symmetry of nature at some high energy scale but be broken by 
non-zero vacuum expectation values (vev's) of scalar fields called flavons.  

In this report, we describe the implications for Large Hadron Collider (LHC) experiments of a 
simplified model of broken gauged flavor symmetry published recently~\cite{Berger:2014gga}.  
The minimal new particle content of the model includes a scalar flavon, a gauge singlet under the SM;  
a heavy fermion $T$ partner of the top quark; and a neutral top-phiilc gauge boson $Z_T$.  The mass 
of the flavon could be at the TeV scale.  We concentrate here on the modifications of Higgs boson 
production and decay properties introduced by this model and on flavon phenomenology at the LHC, 
notably its production cross section and the decay modes $\varphi \rightarrow h h$ and 
$\varphi \rightarrow Z Z$.    

\section{Simplified model of broken flavor symmetry}
\label{sec:model}

A ``minimal" model of broken flavor symmetry was proposed in~ \cite{Grinstein:2010ve}, and we adopted 
this approach as our starting point.  This model has exotic fermion partners of the 
SM quarks, flavor gauge bosons, and two scalar flavon fields $Y_u$ and $Y_d$ for the up-like and down-like quarks.  The large hierarchy between the masses of the SM quarks corresponds to a large hierarchy between the vevs of the flavons which suggests that the flavor symmetry could be broken sequentially~\cite{Feldmann:2009dc}.  If we integrate out the heavy  degrees of freedom associated with the first and second generations, we are left with a simplified flavor symmetry model with a manageable number of beyond-the-SM (BSM) degrees of freedom (a flavon, an exotic fermion, and a massive vector boson) associated with the top-quark and bottom-quark sectors. 
Because the vev of the flavon associated with the bottom-quark is nearly two orders of 
magnitude larger than the vev of the flavon associated with the top-quark, we also integrated out the flavon associated with the bottom-quark.  At the TeV scale, we are left with the effective Lagrangian
\be
\mathcal{L}_{\text{topflavor}}=\lambda\bar Q_L \tilde{H} \Psi_{tR}-\lambda^\prime\bar \Psi_{t} \Phi 
\Psi_{tR}-M\bar \Psi_{t} t_R+{\text{h.c.}} .
\label{eq:ope}
\ee
Here $\Phi$ is a complex flavon associated with the top-quark; $Q_L$ is the SM quark field,  
$\Psi_{t}$ are $\Psi_{tR}$ are the top-partner fermion fields; $H$ is the 
SM Higgs doublet field, and $\tilde H_i\equiv \varepsilon_{ij}H_j$ where {$\varepsilon_{ij}$ is the anti-symmetric 
tensor with $\varepsilon_{12}=1$.  $\lambda$ and $\lambda^\prime$ are dimensionless parameters,  
$M$ is a parameter with the dimensions of mass. 

 After electroweak symmetry breaking and flavor symmetry breaking, 
\bea
H&=&\left(\begin{array}{c}0 \\\displaystyle{\frac{v+\tilde h}{\sqrt 2}}\end{array}\right),\\
\Phi&=&\frac{\tilde \varphi+v_\varphi}{\sqrt2},
\eea
in the unitary gauge, where $v=246$ GeV is the vev of the Higgs field,
$\tilde h$ is the physical degree of freedom of the SM Higgs doublet field, and $\tilde \varphi$
is the physical degree of freedom of the top flavon. 
The mass eigenstates $\left(h,\varphi\right)$ of the scalar fields are 
 linear combinations of $\tilde h$ and $\tilde \varphi$: 
\be
\left(\begin{array}{c}\tilde h \\\tilde \varphi\end{array}\right)=
\left(\begin{array}{cc}\cos\theta_H & \sin\theta_H \\ 
-\sin\theta_H &\cos\theta_H\end{array}\right)
\left(\begin{array}{c}h \\\varphi\end{array}\right),
\ee
where $\theta_H$ is the Higgs-flavon ``mixing" angle.  
The mixing term, even if forbidden artificially at tree-level, will be generated through 
loop corrections.  The deviation of the Higgs field self-interaction strength $\lambda_H$
from its value $\lambda_H^{SM}=m_h^2/v^2$ in the SM can be written as 
\be
\lambda_H\equiv\lambda_H^{SM}+\frac{m_\varphi^2-m_h^2}{v^2}\sin^2\theta_H.
\ee
To simplify notation, we use $c_H \equiv \cos\theta_H$ and $s_H \equiv \sin\theta_H$.  

The influence of the new physics discussed here on the SM electroweak precision observables 
can be described with the oblique parameters $S$, $T$, $U$ \cite{Peskin:1990zt}.  
The contribution from the exotic real scalar boson $\varphi$ to the oblique 
parameters \cite{Barger:2007im} is suppressed by the
mixing angle $\theta_H$.  We showed the one standard deviation
(1$\sigma$), 2$\sigma$ and 3$\sigma$ fit regions in Ref.~\cite{Berger:2014gga}.  
A detailed analysis of $\Delta F=2$ flavor physics observables and of
$B\to X_s\gamma$ in this model was presented in Ref. \cite{Buras:2011wi}.  

\section{Modifications of Higgs boson physics}
\label{sec:higgs}

The interactions between the SM-like Higgs boson and other 
SM particles are different from those in the pure SM.  The differences have  
two origins.  First, there is mixing between the $SU(2)_L$ doublet and the 
flavon.  Second, there is mixing between the SM top-quark and the 
heavy fermion $T$.  We investigated the production cross section and the decay 
properties of the SM-like Higgs boson $h$ in this gauged broken flavor symmetry 
model.  

Gluon fusion is the most important production channel of the SM Higgs boson 
at the LHC.  In the NP model, the interaction between the SM-like Higgs boson 
and the gluon is mediated by both the SM top-quark and the heavy fermion $T$.  
Denoting the SM and the NP $hgg$ interactions as
\be
c_{hgg}^{SM}hG_{\mu\nu}^aG^{\mu\nu,a},~~~
c_{hgg}^{NP}hG_{\mu\nu}^aG^{\mu\nu,a},
\ee
respectively, we obtained 
\bea
\frac{c_{hgg}^{NP}}{c_{hgg}^{SM}}&\to&\frac{\lambda vc_H}{\sqrt2}
\left(\frac{c_Ls_R}{m_t}-\frac{s_Lc_R}{m_T}\right)\nonumber\\
&&-\frac{\lambda^\prime vs_H}{\sqrt2}
\left(\frac{s_Ls_R}{m_t}+\frac{c_Lc_R}{m_T}\right)\nonumber\\
&=& c_H ,
\eea
in the limit of large fermion mass ($m_{t,T}\gg m_h$).  

We showed that the strengths of most of the Higgs coupling vertices are just 
rescaled by a factor $c_H$.  We examined 
the loop induced interactions also.  The $hgg$ and $h\gamma
\gamma$ vertices are rescaled by the factor $c_H$ in the heavy fermion limit.
The $hZ^0\gamma$ vertex deviates from the simple $c_H$ rescaling, but the 
deviation is small.  The Higgs boson decay branching ratios are nearly unchanged 
relative to the SM since every sizable partial width is changed by an overall factor $c_H^2$.

In the model, the inclusive Higgs production cross section is suppressed 
by a factor $c_H^2$, allowed by the LHC data at 7 and 8 TeV.  
The result from a fit of the Higgs boson inclusive cross section $\mu=\sigma/\sigma_{SM}$ 
by the CMS collaboration \cite{CMS-PAS-HIG-13-005} is 
\be
\mu=0.80\pm0.14.
\ee
The result from the ATLAS collaboration 
\be
\mu=1.30\pm0.12({\text{stat}})^{+0.14}_{-0.11}({\text{sys}}).
\ee
would exclude most of the parameter space of the NP model~\cite{ATLAS-CONF-2014-009}.
However, at the 3~$\sigma$ C.L., the region $s_H^2<0.2$ is still allowed.

\section{Flavon phenomenology at the LHC}
\label{sec:lhc}
The flavon is produced dominantly through gluon fusion at the LHC.  When the flavon is heavy, the 
heavy fermion limit is not a good approximation, and, therefore, the naively expected suppression 
factor $s_H^2$ is not valid.  We calculated the flavon production cross 
section using the complete expressions found in Ref.~\cite{Berger:2014gga}.  In contrast to the 
gluon fusion case, the flavon cross sections 
in the VBF and vector boson associated production channels, where loop effects do not play a role, 
are just rescaled by a factor of $s_H^2$ relative to the Higgs boson cross sections.

The flavon has the same decay modes as the SM Higgs boson.  For a light flavon 
($m_\varphi <2m_t$), the partial decay widths of the regular decay channels 
(except $\varphi  \rightarrow gg, \gamma\gamma, \gamma Z$) are all rescaled by a 
factor of $s_H^2$ which will not affect the decay branching ratios of the flavon.  
In addition, the $\varphi \to hh$ decay width is important, and, for a relatively heavy 
flavon ($2m_t<m_\varphi <m_t+m_T$), the $\varphi \to t\bar t$ decay width. 
Since there is no $s_H^2$ suppression in the $\varphi t\bar t$ vertex, the 
flavon will decay into $t\bar t$ with a large branching ratio at small $s_H$
if allowed by phase space.
However, the flavon production cross section is highly suppressed 
by $s_H^2$ in this region.   The signal will be hidden under the SM $t\bar t$ 
background making the signal for the flavon hard to find in this mode.

For a heavy flavon, as noted before, the heavy fermion limit is not a good approximation.  
We calculated the branching 
ratios of the loop-induced processes ($\gamma\gamma, gg, Z^0\gamma$) using the 
exact formula.  For $m_\varphi >160$ GeV, the contribution from the loop induced channels 
is negligibly small, and the most important decay modes are $b\bar b,
t\bar t, W^+W^-, Z^0Z^0$, and $hh$.  
The branching ratios for the dominant decay channels are shown 
in Fig. \ref{fig:sdecaybr}; $\varphi \to hh$ is an important decay channel 
of the flavon, and it might be used for discovery.  Searches at the LHC can 
focus on the SM Higgs-like decay channels ($Z^0Z^0$, $W^+W^-$) and on the 
light Higgs boson pair decay channel. 

%\begin{figure*}[!htb]
\begin{figure}[!htb]
\begin{center}
\includegraphics[scale=0.38,clip]{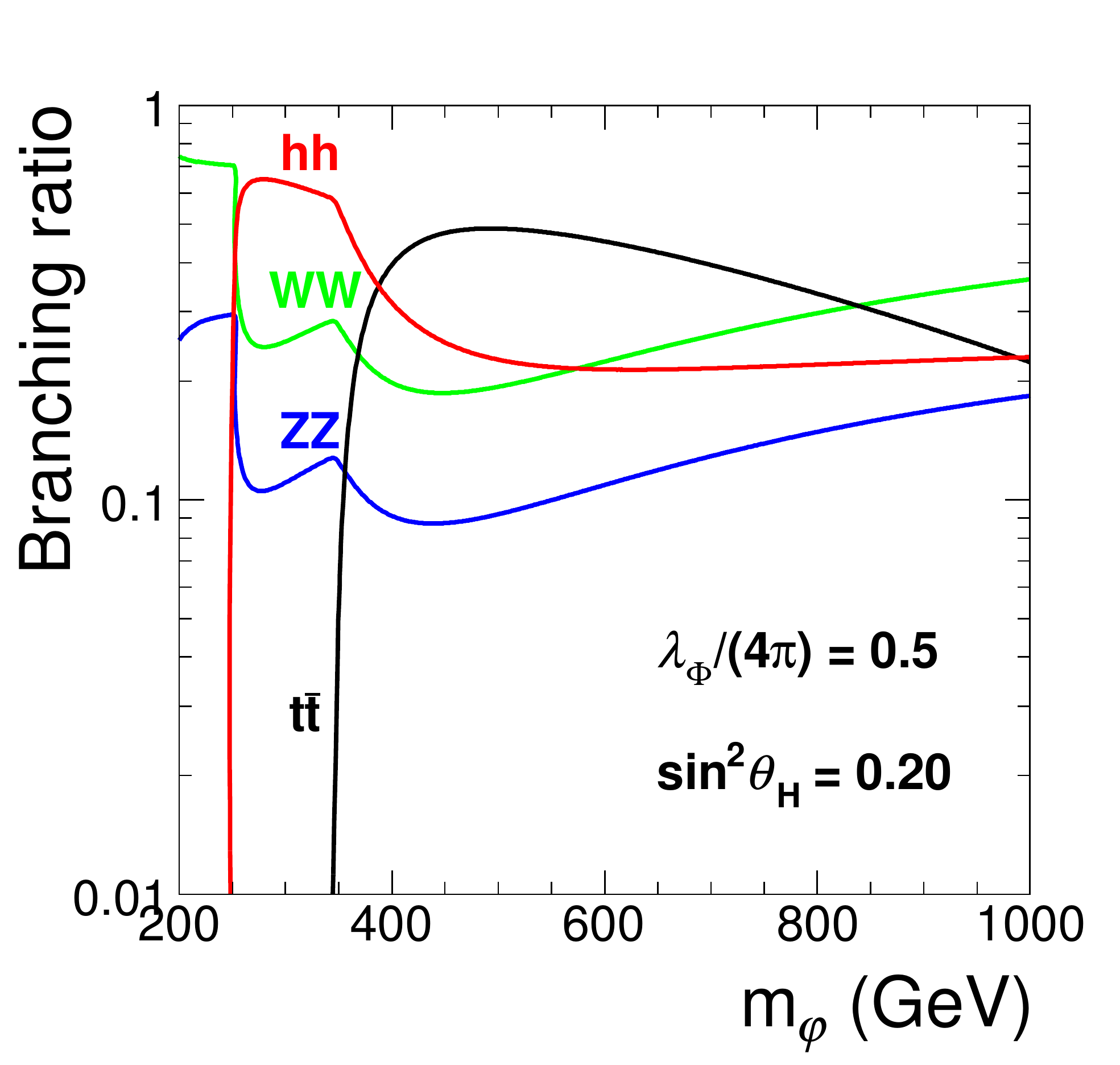}
\caption{
The decay branching ratios of the most important $\varphi$ decay channels.
\label{fig:sdecaybr} }
%\end{figure*}
\end{center}
\end{figure}

At 7 and 8 TeV, the strongest limit on the flavon is provided by the 
$Z^0Z^0\to 2\ell2\ell'$ channel in heavy SM Higgs boson 
searches \cite{ATLAS-CONF-2013-013,Chatrchyan:2013mxa}.  The CMS 
collaboration also investigated the $hh$ channel \cite{CMS-PAS-HIG-13-025,
CMS-PAS-HIG-13-032}.  Limits from the 7 and 8 TeV LHC data are presented 
in Ref.~\cite{Berger:2014gga}.  

\subsection{Flavon searches at 14 TeV}
We investigated the possibility of discovering a flavon at 14 TeV with 100 fb$^{-1}$ integrated 
luminosity.  There are simulations of the $Z^0Z^0$ channel by the ATLAS collaboration
\cite{ATLAS-PHYS-PUB-2013-016} and the CMS collaboration
\cite{CMS-PAS-FTR-13-024} for this channel.   We rescaled their upper bounds to 
100 fb$^{-1}$ integrated luminosity.  When $m_\varphi  <2m_h$, the $Z^0Z^0$ channel can 
provide a very strong constraint on the NP model 
(e.g., $s_H^2$ greater than $\sim 0.08$ is excluded).  
When $m_\varphi >2m_h$, the constraint on  
$s_H^2$ is at $\mathcal{O}\left(10^{-1}\right)$. In this region of $s_H^2$,
the $hh$ channel 
will be the dominant decay channel of $\varphi$.
\begin{figure*}[!htb]
\begin{center}
\includegraphics[scale=0.82,clip]{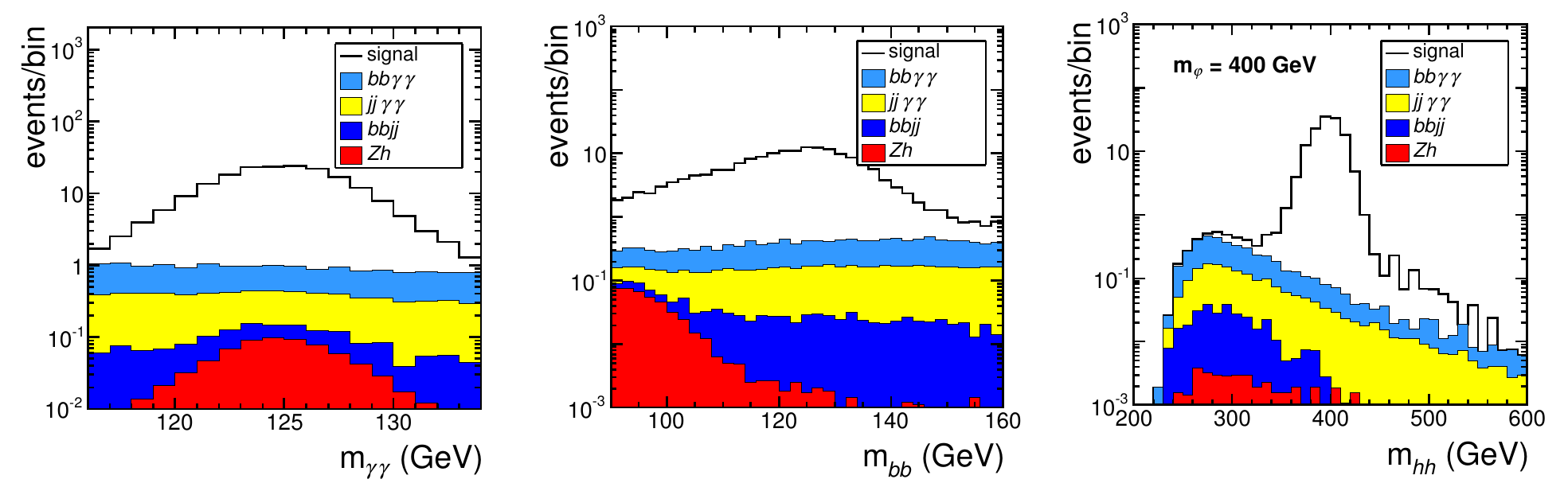}
\caption{The reconstructed diphoton, $b\bar b$, and $hh$ mass distributions 
are shown at 14 TeV with 100 fb$^{-1}$ integrated luminosity.
We choose $m_\varphi =$400 GeV. The total cross section is rescaled to 
the value of the SM like Higgs boson with the same mass. The decay branching ratio
${\text{Br}}\left(\varphi \to hh\right)$ is set to be 100\%. Note that the horizontal scale 
differs in the three distributions.
\label{fig:s300} }
\end{center}
\end{figure*}
We focus on the $b\bar b\gamma\gamma$ channel and present the results of our 
detailed simulation of the signal and backgrounds. 

We generated the signal and background events at the parton level using 
MadGraph5 \cite{Alwall:2011uj,Alwall:2014hca}.   
For signal events, we generated $pp\to \varphi +{\text{n}}j$ 
to n=1. 
All of the parton level signal and the background events were showered using 
Pythia6.4 \cite{Sjostrand:2006za}. The MLM matching scheme \cite{Mangano:2006rw} 
was used to avoid double counting.  Detector effects
were mimicked with PGS4 \cite{pgs}.  Jets were defined in the events with the 
anti-$k_T$ algorithm, with $R=0.4$.  
The cross section for the signal was normalized to the NNLO SM-like Higgs boson 
cross section suggested by the LHC Higgs Cross Section Working 
Group~\cite{Dittmaier:2011ti} multiplied by the rescaling factor from the 
$gg\varphi$ vertex. 

There are several irreducible SM backgrounds
\bea
pp&\to& b\bar b \gamma\gamma,\nonumber\\
pp&\to& Z^0h \to b\bar b \gamma\gamma,\nonumber\\
pp&\to& Z^0\gamma\gamma\to b\bar b \gamma\gamma,
\eea
and reducible SM backgrounds
\bea
pp&\to& b\bar b jj~\left(j\to \gamma\right),\nonumber\\
pp&\to& jj\gamma\gamma~\left(j\to b\right),\nonumber\\
pp&\to& t\bar t\to bjj\bar bjj~\left(j\to \gamma\right),\nonumber\\
pp&\to& t\bar t h\to b\ell^+ \nu\bar b\ell^- \bar\nu\gamma\gamma
~\left(\ell^\pm~{\text{missed}}\right).
\eea
Next-to-leading order contribution were 
included where available, with $K$ factors if necessary.  

We required the events to have at least 
two hard isolated photons in the central region, 
\be
p_T^\gamma>20 {\text{GeV}},~~\left|\eta^\gamma\right|<2.0,
\ee
and no hard jet or charged lepton in the $\Delta R=0.4$ region around the
photon.  We demanded at least two hard $b$-tagged jets with
\be
p_T^j>40~ {\text{GeV}},~~\left|\eta^j\right|<2.0.
\ee
The average $b$-tagging efficiency was reweighted to 70\% \cite{ATL-PHYS-PUB-2013-009} 
in the analysis.  To suppress the SM $t\bar th$ background, we rejected events which 
contained hard isolated charged leptons 
and events with large missing transverse energy $\met>30~ {\text{GeV}}$.

Signal events were required to satisfy hard cuts designed for the Higgs boson pair 
signal, with cuts on the leading and sub-leading photon:  
\be
\left|m_{\gamma\gamma}-125.4 ~{\text{GeV}}\right|<\Delta m_{h,{\text{cut}}}^{\gamma\gamma}.
\ee
The transverse momentum of the leading and of the sub-leading photon had to satisfy
\be
p_T^{\gamma_1}>p_{T,{\text{cut}}}^{\gamma_1},~~p_T^{\gamma_2}>p_{T,{\text{cut}}}^{\gamma_2}.
\ee
The leading and sub-leading $b$-tagged jets had to satisfy
\be
\left|m_{bb}-125.4 ~{\text{GeV}}\right|<\Delta m_{h,{\text{cut}}}^{bb}.
\ee
$\delta\phi_{\gamma b}$, the smallest of $\Delta\phi_{\gamma_1b_1},\Delta\phi_{\gamma_1b_2},\Delta\phi_{\gamma_2b_1},
\Delta\phi_{\gamma_2b_2}$ (the differences between the azimuthal angles of the objects) 
was required to be less than $\Delta\phi_{\gamma b}$.

We reconstructed the invariant mass peak of the flavon after including the energy 
resolution. In Fig. \ref{fig:s300}, we show the results for the reconstructed diphoton, 
$b\bar b$, and $hh$ mass distributions using events which satisfy all the cuts.
The resonance signal is clear.  
The dominant background is $b\bar b\gamma\gamma$,  
and the other backgrounds are numerically small. The diphoton 
resonance is very clear; the $b\bar b$ peak is wide 
owing to the larger energy smearing of jets.  

After a scan over the mass of the flavon, we derived expected limits from the search 
for the $\varphi \to hh\to b\bar b\gamma\gamma$ signal at 14 TeV with 
100 fb$^{-1}$.  These limits can be translated into a constraint on the NP parameters. 
\begin{figure}[!htb]
\begin{center}
\includegraphics[scale=0.38,clip]{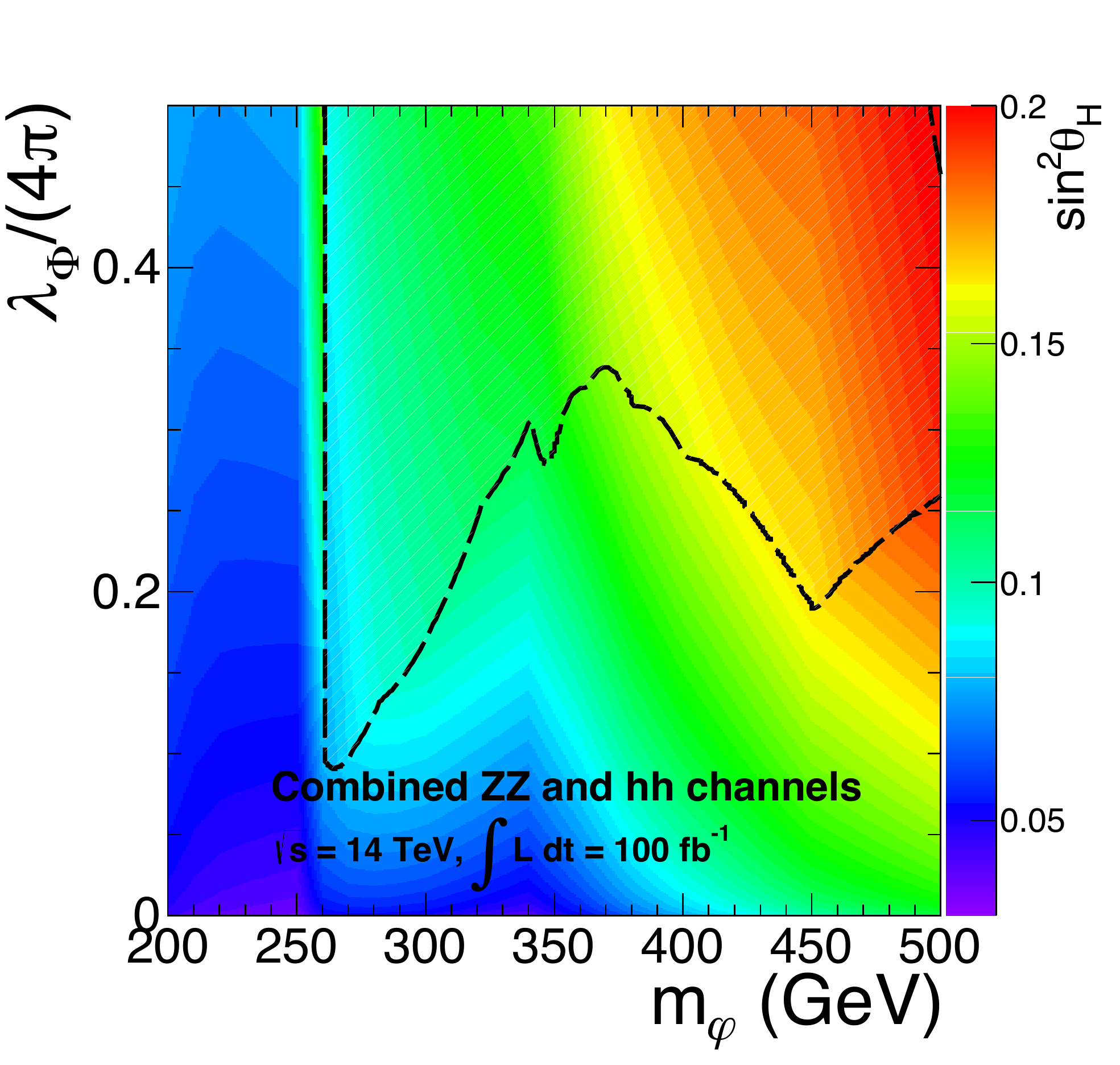}
\caption{Combination of the 
2$\sigma$ exclusion region of $s_H^2$
results for the $\varphi \to Z^0Z^0 \to 2\ell2\ell^\prime$ search
and the $\varphi \to hh \to b\bar b\gamma
\gamma$ search.  In the upper part of the figure, in the irregularly 
shaped region above the broad-dashed line,  
the $\varphi \to hh \to b \bar b \gamma \gamma$ 
search yields a stronger constraint.
\label{fig:14hh} }
\end{center}
\end{figure}
\begin{figure}[!htb]
\begin{center}
\includegraphics[scale=0.38,clip]{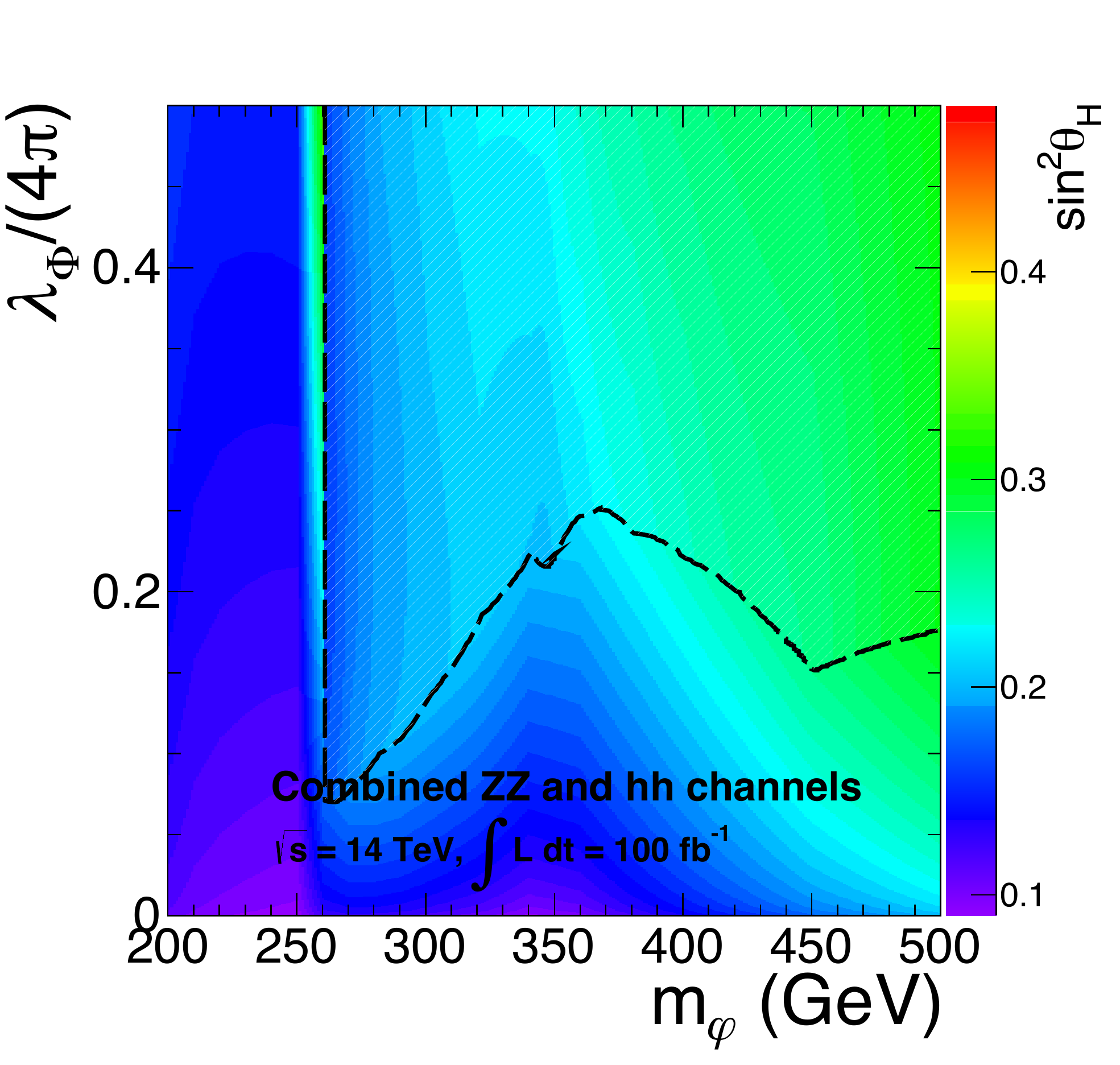}
\caption{The 5$\sigma$ discovery significance of the required value of $s_H^2$ 
from a combination of the $\varphi \to Z^0Z^0 \to 2\ell2\ell^\prime$ search
and the $\varphi \to hh \to b\bar b\gamma
\gamma$ search.  In the upper part of the figure, in the irregularly 
shaped region above the broad-dashed line,  
the $\varphi \to hh \to b \bar b \gamma \gamma$ process is more sensitive to the NP model.}
\label{fig:14hh5} 
\end{center}
\end{figure}
Combining the results from the SM-like heavy Higgs boson search 
for $\varphi \rightarrow Z^0Z^0$ and the $\varphi \to hh\to b\bar b \gamma\gamma$ search 
at 14 TeV with 100 fb$^{-1}$ integrated luminosity,
we show the constraint on $s_H^2$ in Fig. \ref{fig:14hh}.  The combined 5$\sigma$ discovery 
significance is shown in Fig. \ref{fig:14hh5}. 
In FIG. \ref{fig:14hh}, the search for the 
$\varphi \to hh\to b\bar b \gamma\gamma$ signal gives a stronger constraint 
in the cross-hatched region.  This signal can give 
a stronger constraint in the large $\lambda_\Phi$ 
region when the $\varphi\to hh$ channel opens.
This figure shows that a strong constraint on 
the neutral scalar $\varphi$ can be obtained with a combination of the two channels.

\section{Summary}
\label{sec:con}
A model of physics beyond the SM has been proposed in 
which there is a new scalar, a flavon,  a new heavy fermion associated with 
the SM top-quark, and a new neutral flavor gauge boson. 
This model arises as the low-energy limit of a theory of gauged 
flavor symmetry with an inverted hierarchy, giving a simplified 
model with spontaneous breaking of flavor symmetry.  
The flavon mixes with the SM 
Higgs boson, and the heavy fermion alters the production and decay properties of 
the Higgs boson at the LHC, all in ways that are consistent with data at current 
levels of precision. The flavon and the heavy fermion might appear at the hundreds of 
GeV to the TeV scale. There is a sizable allowed parameter space in which existing constraints
from electroweak precision observables and flavor physics are satisfied.

In this NP model, the production cross section of the SM-like Higgs boson at the LHC
is suppressed by a factor $\cos^2 \theta_H$, where $\theta_H$ is the mixing angle between 
the Higgs boson and the flavon.  However, neither mixing nor the triangle loop from the heavy 
fermion change the Higgs boson decay branching ratios significantly.  
The possibility to search for the flavon at the LHC was explored in detail through  
its decays to a SM Higgs-pair, $\varphi \rightarrow hh$, as well as through the SM Higgs 
boson decay modes.  At 7 and 8 TeV at the
LHC, the $Z^0Z^0$ channel will give a stronger constraint than $\varphi \rightarrow h h$
owing to limitations of integrated luminosity.  At 14 TeV with 100 fb$^{-1}$ integrated luminosity, 
the small mixing region can be reached where the $\varphi \rightarrow hh$ signal is 
important for discovery.  The flavon can be produced singly at the LHC.  If it decays into 
the $hh$ final state with a sizable decay branching ratio, the $hh$ cross section will be 
enhanced significantly by this resonance effect.  

%\begin
{\bf Acknowledgments.} The work of E. L. Berger at Argonne is supported 
in part by the U.S. DOE under Contract No. DE-AC02-06CH11357.
H. Zhang has been supported by the U.S.  DOE under Contracts 
No. DE-FG02-91ER40618 and DE-SC0011702.  
We warmly acknowledge the essential contributions of 
Steven Giddings, UC Santa Barbara, and Haichen Wang, 
LBNL, Berkeley, to the research summarized in this brief report.  



%% References

%\nocite{*}
\bibliographystyle{elsarticle-num}
\bibliography{BergerICHEP2014}






% \end{thebibliography}

\end{document}